\documentclass[12pt]{article}
\usepackage{amsmath}
\usepackage{amssymb}
\begin{document}
\vskip 2cm
\begin{center}
{\sf {\Large A Free  $\mathcal{N }= 2$ Supersymmetric System: Novel Symmetries 
}}

\vskip 2.0cm

{\sf S. Krishna$^{(a)}$, R. P. Malik$^{(a,b, c)}$}

\hskip 1cm $^{(a)}$ {\it Physics Department, Centre of Advanced Studies,}\vskip .06cm
\hskip 1.5cm {\it Banaras Hindu University (BHU), Varanasi - 221 005,  India}\vskip .06cm
\hskip 1cm $^{(b)}$ {\it DST-CIMS, BHU, Varanasi - 221 005, (U.P.), India}\vskip .06cm
\hskip 1cm $^{(c)}$ {\it AS-ICTP, Strada Costiera-11, I-34014,
Trieste, Italy}\vskip .05cm
\hskip 1.5cm {\small {\sf {E-mails: skrishna.bhu@gmail.com;  rpmalik1995@gmail.com}}}

\end{center}

\vskip 1cm

\noindent
{\bf Abstract:} 
We  discuss a set of novel discrete symmetries of a free $\mathcal{N} = 2$ supersymmetric (SUSY)
quantum mechanical system which is the limiting case of a widely-studied interacting SUSY model of a 
charged particle constrained to move on a sphere in the background of a Dirac magnetic monopole. The {\it usual} 
continuous symmetries of this model provide the physical realization of the de Rham cohomological operators of 
differential geometry. The interplay between the {\it novel} 
discrete symmetries and {\it usual} continuous symmetries leads to the physical realization of relationship between the (co-)exterior derivatives of differential geometry.   We have also  exploited the supervariable 
approach to derive the nilpotent $\mathcal{N}  =2$ SUSY symmetries of the theory and provided the geometrical
origin and interpretation for the nilpotency property. 
Ultimately,  our present study  (based on innate symmetries) proves that our {\it free}
$\mathcal{N} = 2$ SUSY example is a tractable model for the Hodge theory.




\newpage
\section{Introduction}

It is a well-known fact that {\it three} out of {\it four} fundamental interactions of nature are theoretically  described 
within the framework of gauge theories which are characterized by the existence of  local
gauge symmetries at the {\it classical} level (see, e.g. [1]). A very spacial class of gauge theories is endowed 
with the dual-gauge symmetries, too. For instance, it has been shown recently that any arbitrary Abelian $p$-form
($p = 1, 2, 3...$) gauge theory would be always endowed with the (dual-)gauge symmetries 
 in $D = 2p$ dimensions of spacetime [2,3,4]. As a consequence, such theories have been shown,
within the framework of Becchi-Rouet-Stora-Tyutin  (BRST) formalism, to respect (at the {\it quantum} level)
the (anti-)BRST and (anti-)dual-BRST
symmetries in the Feynman gauge  (see, e.g. [2-5] for details).
Mathematically, such field theoretic models have been shown to present a set of  tractable  examples for the Hodge
theory (see, e.g. [3-9]) because the continuous and discrete symmetries of such a class of theories provide
the physical realizations of the de Rham cohomological operators of differential geometry.

In a recent set of papers (see, e.g. [10-12]), a collection of $\mathcal{N} = 2$ SUSY quantum mechanical 
models  (QMM) have {\it also} been shown to represent the models for the Hodge theory where the symmetries 
(and their generators) play an important role. The central purpose  of our present endeavor  is to show that the free $\mathcal{N} = 2$ SUSY QMM (which is the limiting case of the physically interesting model of a charged particle,
constrained  to move on a sphere, in the background of a Dirac magnetic monopole) {\it also} presents  an example for the Hodge theory. This is essential {\it first} modest step for us if we wish to prove, in a systematic 
manner, that its $\mathcal{N} = 2$ and $\mathcal{N} = 4$ {\it interacting} versions are also models for the Hodge theory.

There are a few continuous symmetries (and their conserved charges) and discrete symmetries 
that are urgently needed to prove a field theoretic model and/or a SUSY QMM 
to be an example for the Hodge theory. Such studies are physically relevant  because, taking the help 
of this kind of investigations, we have 
proven that the (1 + 1)-dimensional (2D) (non-)Abelian gauge theories (without any interaction with the
matter fields) are perfect examples of a {\it new} class of topological field theory (TFT) which capture a few 
key aspects of the Witten-type TFT and some of the salient features of Schwartz-type TFT (see, e.g. [13]). An interacting   system of 2D photon and Dirac fields has also been shown to be a model for the Hodge theory where the topological gauge field couples with the matter fields [9,14]. Similar is the case with our very recent works on
the modified versions of the 2D anomalous  gauge theory and Proca theory where matter and gauge fields are present together [15,16]. 

The free $\mathcal{N} = 2$ SUSY system under consideration is interesting in its
own right because, since the seminal work by Dirac [17],  the system of charged particle and magnetic monopole
has been studied from different angles  due to its rich mathematical and physical structures (see, e.g. [18,19]).
The $\mathcal{N} = 2$  (and its generalization to $\mathcal{N} = 4$)  
superfield formulations have been carried out in [20,21] where the quantum mechanical Lagrangian for the above physical system has been obtained by adopting the $CP^{(1)}$ model approach so that there is {\it no singularity} in the monopole
interaction. In our present investigation, we take the {\it free} $\mathcal{N} = 2$ SUSY version of the Lagrangian obtained in [20] and show that it provides the physical model for a Hodge theory
due to its innate symmetries.


The material of our present paper is organized in the following fashion. In Sec. 2, we recapitulate the bare 
essentials of the $\mathcal{N} = 2$  nilpotent  ($s_1^2 =  s_2^2 =0$)
 SUSY transformations ($s_1$ and $s_2$) and a bosonic symmetry $s_\omega = \{s_1, \, s_2\}$. 
 Our Sec. 3 is devoted to the discussion of a set of discrete symmetry transformations.
In Sec. 4, we lay emphasis on the algebraic structures of the symmetry transformations and corresponding conserved charges.
The $\mathcal{N} = 2$ SUSY continuous nilpotent symmetry transformations (i.e. $s_1$ and $s_2$) 
are derived by exploiting the SUSY invariant restrictions in Sec. 6. Finally, we make 
some concluding remarks 
in Sec. 7.

\section{Preliminaries: usual  continuous symmetries  }

We begin with the $\mathcal{N} = 2$ SUSY invariant Lagrangian (derived by exploiting the standard technique
of the superspace formalism) for the case of a charged particle moving on a sphere in the background of a
Dirac magnetic monopole as (see, e.g. [20] for details) 
\begin{eqnarray}
L &=& 2\, { D}_t {\bar z} \cdot D_t z + \frac{i}{2}\, \left[\bar\psi \cdot D_t\psi
- { D}_t {\bar\psi}\cdot \psi \right] - 2\, g\, a,
\end{eqnarray} 
where $D_t z = (\partial_t - i\,a )\,z \equiv (\dot z - i\, a z),\, { D}_t \bar z= (\partial_t 
+ i\,a )\,\bar z \equiv \dot{\bar z} + i\, a\, \bar z, \, D_t\psi=  (\partial_t - i\,a )\,\psi
\equiv \dot \psi - i\, a\, \psi,\,
 { D}_t \bar\psi = (\partial_t + i\,a )\,\bar\psi \equiv \dot{\bar \psi} + i\, a\, \bar\psi$
are the $U(1)$ covariant derivatives under the $CP^{(1)}$ model approach  with the real ``gauge" variable 
$a$ and $\partial_t = d/dt$ is the derivative  w.r.t. the evolution parameter $t$. Here the electric 
charge $e$ of the particle (with mass $m = 1$) is taken to be $e = - 1$ and the magnetic
 charge on the monopole is denoted by $g$.

We concentrate on the {\it free} case of the above Lagrangian where $a = 0$. This leads to the following [20] 
 \begin{eqnarray}
L_0 &=& 2\,\dot {\bar z}\cdot \dot z + \frac{i}{2}\,\left(\bar\psi\cdot\dot\psi
- \dot{\bar\psi}\cdot\psi \right),
\end{eqnarray}
where $\bar z \cdot z =  |z_1|^2 + |z_2|^2$ because we have taken $\bar z = ({\bar z}_1 \;\; {\bar z}_2)$
and $z = (z_1 \;\; z_2)^T$ 
as complex variables. Similar is the case with $\bar\psi \cdot \psi = \bar \psi_1\, \psi_1 
+ \bar\psi_2 \,\psi_2$ because $\bar\psi$ and $\psi$ are {\it independent} fermionic variables with
$\psi\cdot\psi = 0$ and  $\bar\psi\cdot \bar\psi = 0$ where $\psi\cdot\psi \equiv \psi^T \cdot \psi = \psi_1^2 +
\psi_2^2 = 0$, etc.

The continuous and nilpotent ($s_1^2 = s_2^2 = 0$) $\mathcal{N} = 2$ SUSY  symmetry 
transformations of the above free Lagrangian $L_0$ are as follows:
\begin{eqnarray}
&& s_1 z = \frac{\psi}{\sqrt 2}, \quad s_1 \psi = 0, \quad s_1 \bar\psi 
= \frac{2\,i\,  \dot {\bar z}}{\sqrt 2}, \quad s_1 \bar z = 0, \nonumber\\
&& s_2 \bar z =  \frac{\bar\psi}{\sqrt 2}, \quad s_2 \bar\psi = 0, \quad 
s_2 \psi = \frac{2 \, i\,   \dot z}{\sqrt 2}, \quad  s_2 z = 0,
\end{eqnarray}
because the Lagrangian $L_0$ transforms to 
\begin{eqnarray}
s_1\, L_0 = \frac{d}{dt} \left(\frac{\dot {\bar z} \cdot \psi}{\sqrt 2}\right),
 \qquad \qquad s_2\, L_0 = \frac{d}{dt} \left(\frac{\bar\psi \cdot\dot z}{\sqrt 2} \right).
\end{eqnarray}
As a consequence,  the action integral $S = \int dt\, L_0$ remains invariant.
It is easy to check that the generators  
of the above transformations are the  conserved  charges:
\begin{eqnarray}
Q = \frac{2\, \dot{\bar z} \cdot\psi}{\sqrt 2} \equiv \frac{\Pi_z\cdot \psi}{\sqrt 2},
 \qquad\qquad \bar Q = \frac{2\, \bar\psi \cdot \dot z}{\sqrt 2} 
\equiv \frac{\bar\psi\cdot \Pi_{\bar z}}{\sqrt 2},
\end{eqnarray}
where the canonical conjugate momenta $\Pi_z$ and $\Pi_{\bar z}$ are w.r.t. variables $z$ and $\bar z$.
Similarly, the conjugate momenta w.r.t. $\psi$ and $\bar\psi$ in our theory are: $\Pi_\psi = - (i/2) \, \bar \psi$
and $\Pi_{\bar\psi} = - (i/2) \,  \psi$ where the convention of the left-derivative w.r.t. the fermionic variables 
$\psi$ and $\bar\psi$ has been adopted. The conserved charges $Q$ and $\bar Q$ are the generators for $s_1$ and
$s_2$ as can be seen from the following relationships:
\begin{eqnarray}
s_r \Phi = \pm \,i\; [\Phi,\, Q_r]_\pm \qquad \qquad r = 1, 2 \quad(Q_1 = Q, \; Q_2 = \bar Q),
\end{eqnarray} 
where $\Phi = z, \bar z, \psi, \bar\psi$ is the generic variable of our present  theory and subscript ($\pm$) 
on the square bracket corresponds to the (anti)commutator for the generic variable 
$\Phi$ being  (fermionic)bosonic in nature. 

The anticommutator of $s_1$ and $s_2$ (i.e. $s_\omega = \{s_1, \; s_2\}$) generates a
bosonic symmetry in the theory, namely;
\begin{eqnarray}
s_\omega z = \dot z, \quad\qquad s_\omega \bar z = \dot{\bar z}, \quad\qquad s_\omega \psi = \dot\psi,
\quad\qquad s_\omega \bar\psi = \dot{\bar \psi},
\end{eqnarray}
modulo a factor of $i$. It is obvious that the generator of this time-translation is nothing but  the Hamiltonian 
of our present free $\mathcal{N}  =2$ SUSY theory. The explicit expression for the Hamiltonian (of our 
free $\mathcal{N}  =2$ SUSY system) is
\begin{eqnarray}
H = 2 \,\dot{\bar z}\cdot \dot z \equiv \frac{\Pi_z \cdot \Pi_{\bar z}}{2},
\end{eqnarray}
where there is no potential (i.e. interaction) term. 

\section{Novel discrete symmetries}

Under the following discrete transformations\footnote{
In these discrete symmetry transformations, 
we have suppressed the explicit notations for the transpose operations on the dynamical variables
$z, \bar z, \psi, \bar \psi$ of our SUSY quantum mechanical theory.}:
\begin{eqnarray}
&&  z \rightarrow  - \,\bar z,\;\, \quad \bar z \rightarrow   -\, z, \;\quad\,
\psi \rightarrow  -\, \bar \psi, \;\,\quad \bar\psi \rightarrow  
+ \, \psi, \qquad t \rightarrow -\, t,  \nonumber\\
&& z \rightarrow  \pm \, i \,\bar z, \,\quad \bar z \rightarrow   \mp\, i\,  z, \,\quad
\psi \rightarrow  \pm \, i\, \bar \psi, 
 \bar\psi \rightarrow  \pm\, i \, \psi, \qquad t 
\rightarrow -\, t,
\end{eqnarray}
the Lagrangian $L_0$ remains invariant. We note that there is a time-reversal (i.e. $t \rightarrow - t$)
symmetry in the theory which implies, e.g., $z(t)\rightarrow z(-t) = \pm\, i \,{\bar z}^T$, etc., in the latter 
transformations of (9). The above set of discrete symmetry transformations are the {\it novel} 
useful symmetries because they establish a set of connections  between the nilpotent $\mathcal{N} = 2$ symmetry transformations
$s_1$ and $s_2$ as: 
\begin{eqnarray}
&& s_2 \Phi = \pm\, * s_1\, *\, \Phi, \qquad\qquad
 s_1 \Phi = \mp\, \,*\, s_2\, * \,\Phi, 
\end{eqnarray} 
where ($*$) is nothing but the novel discrete symmetry transformations (9) for the generic variable 
$\Phi = z, \bar z, \psi, \bar\psi$. 

For the duality-invariant theories [22], the ($\pm$) signs in (10) are governed by two successive ($*$)
operations on the generic variable $\Phi$, namely;
\begin{eqnarray}
*\,(*\, \Phi) = \pm\,  \Phi, \qquad\qquad\qquad \Phi = z, \bar z, \psi, \bar\psi.
\end{eqnarray} 
In our case, it can be explicitly checked that  
\begin{eqnarray}
&& *\; (\; *\;   \Phi_1) = \; +\; \Phi_1,  \qquad \qquad \Phi_1 = z, \, \bar z, \nonumber\\
&& *\; (\; *\;   \Phi_2) = \; -\; \Phi_2,  \qquad \qquad \Phi_2 = \; \psi , \; \bar \psi.
\end{eqnarray}
The ($\pm$) signs in $ s_2 \Phi = \pm\, * \,s_1\, *\, \Phi$ (cf. (10)) are the same as those given in (11).
However, it is the reverse signature that is true for  $s_1 \Phi = \mp\, *\, s_2\, * \,\Phi$
{\it vis-\`a-vis} Eq. (11).

It is  interesting to note that, under the following set of discrete  transformations:
\begin{eqnarray}
&& t\rightarrow t, \;\;\;\quad z \rightarrow  \pm \, i \,\bar z, \;\;\;\quad \bar z \rightarrow   \mp\, i\,  z, \;\;\;\quad
\psi \rightarrow  \pm \, i\, \bar \psi, \quad \bar\psi \rightarrow  \mp\, i \, \psi, \nonumber\\ 
&& t\rightarrow t,\; \qquad z \rightarrow  \pm \, i \,\bar z,\; \qquad \bar z \rightarrow   \mp\, i\,  z, 
\qquad \; \psi \rightarrow    \bar \psi,\, \;\qquad \bar\psi \rightarrow   \psi, \nonumber\\ 
&& t \rightarrow t, \;\qquad z \rightarrow   \,\bar z, \qquad\qquad  \bar z \rightarrow   \, z, \qquad \;
 \psi \rightarrow  \, \bar \psi, \quad\qquad \bar\psi \rightarrow  
 \, \psi, 
\end{eqnarray}
the Lagrangian remains invariant. We note that there is {\it no} time-reversal 
symmetry in the above transformations (i.e. $t\rightarrow t$). Further, it is elementary  to note 
 that $*\,(*\, \Phi) = + \,  \Phi$
 for the generic variable $\Phi = z, \bar z, \psi, \bar\psi$.
However, it can be explicitly checked that the relationships:
\begin{eqnarray}
&& s_2 \Phi = + \; * s_1\; *\; \Phi, \qquad\qquad
 s_1 \Phi \ne - \, \;*\; s_2\; * \;\Phi, 
\end{eqnarray} 
are {\it true} for the top and bottom symmetry transformations in (13).
Thus, we note that the reciprocal relationship (i.e. $s_1 \Phi = - \, \;*\; s_2\; * \;\Phi$) is
{\it not} satisfied. It is remarkable to note that {\it even} the first relationship of (14)
is {\it not} satisfied by the discrete  transformations pointed out in the middle of (13). 
Thus, according to the rules laid down in [22] for developing the 
duality-invariant theories,
the discrete transformations (13) are {\it not} useful to us.

 \section{Algebraic structures}

 As we have seen, there are {\it three} continuous symmetries (i.e. $s_1, s_2, s_\omega$)
in the theory. The continuous symmetries, in their operator form, satisfy 
\begin{eqnarray}
&& s^2_1  = 0,\,\quad s^2_2  = 0, \,\quad \{s_1,\, s_2\} = s_\omega = (s_1 + s_2)^2, \nonumber\\
&& \big[s_\omega,\, s_1 \big] = 0,\quad [ s_\omega,\, s_2 ] = 0,\quad \{s_1,\,s_2\} \ne 0,
\end{eqnarray}
when they operate on the generic field $\Phi = z, \bar z, \psi, \bar\psi$ of the theory.
The above algebra is reminiscent of the algebra obeyed by the de Rham cohomological operators  
of differential geometry (see, e.g. [23,24]), namely;
\begin{eqnarray}
&&d^2 = 0,\qquad \delta^2 = 0, \qquad \{d,\, \delta\} = \Delta = (d + \delta)^2,\nonumber\\
&&\big[\Delta,\, d \big] = 0, \qquad \big[\Delta,\, \delta \big] = 0, \qquad \{d,\, \delta\} \ne 0, 
\end{eqnarray} 
where $d$ (with $d^2 = 0$) in the exterior derivative  $\delta$ (with $\delta^2 = 0$) is the (co-)exterior derivative
 and  $\Delta = (d + \delta)^2$ is the Laplacian  operator.

In an exactly similar fashion, it will be noted that the conserved charges ($Q, \bar Q, H$)
also obey 
the following 
algebra
\begin{eqnarray}
&& Q^2 = {\bar Q}^2 = 0, \quad \{Q, \, \bar Q\} = H, \quad
 \big[H,\, Q\big] = 0, \quad [H,\, \bar Q] = 0, \quad H = (Q + \bar Q)^2,
\end{eqnarray} 
in the case of our present theory. The latter two entries (i.e. $ [H,\, Q] = 0,  \; [H,\, \bar Q] = 0$)
are nothing  but the conservation law for the charges $Q$ and $\bar Q$ (which can be  checked easily 
by {\it either} using directly the Euler-Lagrange equations of motion {\it or} the basic brackets
$[z, \, \Pi_z] = [\bar z, \, \Pi_{\bar z} ]=i, \; \{\psi,\, \bar\psi\}= + 1$). 
The algebra in (17) is one of simplest forms [25] of the $\mathcal{N} =2$ SUSY algebra $sl(1/1)$. 

A close look at (15), (16) and (17) demonstrates that, at the algebraic level,
all these equations are equivalent.
However, we have still not been able to provide the physical realization of the very important
relationship $\delta = \pm\, *d*$ that exists between the (co-) exterior derivatives ($\delta)d$ of differential 
geometry. In this connection, it is pertinent to point out that the relationship (10) (cf. Sec. 3)  provides
the physical realization of $\delta = \pm\, *d*$  and $d = \mp\, *\delta*$ in terms of the innate 
continuous and discrete symmetries
of our present theory.  

\section{Towards cohomological aspects}

We have noted in the previous section that $H$ is the 
Casimir operator
of the algebra (17). Thus, it is clear that $H Q = QH$ and $H\bar Q=  \bar Q H$
imply   that $Q H^{-1} = H^{-1} Q$ and $\bar Q H^{-1}=  {H}^{-1} \bar Q$.
Using (17), it can be seen that  
\begin{eqnarray}
&& \left[\frac{Q\, \bar Q}{H}\, , \; Q\right] = +\, Q, \qquad 
 \left[\frac{Q\, \bar Q}{H}\, , \; \bar Q\right] = -\,\bar Q,\nonumber\\ &&
 \left[\frac{\bar Q\,  Q}{H}\, , \; \bar Q\right] = +\, \bar Q, \qquad
  \left[\frac{\bar Q\,  Q}{H}\, , \; Q\right] = -\, Q.
\end{eqnarray} 
As a consequence of the above equation, it is evident that if we define the eigenvalue equation: 
$({Q\, \bar Q}/{H})\, |\psi>_q = q\, |\psi>_q$ for a state $|\psi>_q$
in the quantum Hilbert space of states, then, we have the validity of the following 
\begin{eqnarray}
&&\left(\frac{Q\, \bar Q}{H}\right)\, Q\,\left|\psi \right>_q = (q + 1)\, Q\, \left|\psi \right>_q, \nonumber\\
&& \left(\frac{Q\, \bar Q}{H}\right)\, \bar Q\,\left|\psi \right>_q 
= (q - 1)\, \bar Q\, \left|\psi \right>_q, \nonumber\\
 && \left(\frac{Q\, \bar Q}{H}\right)\, H\,\left|\psi \right>_q = q \, H\, \left|\psi \right>_q, 
\end{eqnarray}
where $q$ is the eigenvalue of state $|\psi>_q$ w.r.t. the hermitian  
operator $ ({Q\, \bar Q}/{H})$. This observation implies that
$q$ is a real number. A close look at (19) demonstrates that $Q\,|\psi >_q$,  
$\bar Q\,|\psi >_q $  and $H\,|\psi>_q $ have the eigenvalues $(q 
+ 1), (q - 1)$ and $q$, respectively, w.r.t. the operator $ ({Q\, \bar Q}/{H})$
which is a physical operator.

The above observation  in (19) establishes a connection between the set of conserved charges 
$(Q, \bar Q, H)$ and the set of de Rham cohomological  operators ($d, \delta, \Delta$)
because, as we know,  the operation of $d$ on a differential form of degree $q$, raises the degree of the 
form by one (i.e. $d\, f^{(q)} \sim  f^{(q + 1)}$). On the contrary, the action of $\delta$,
on a $q$-form, lowers   the degree of form by one (i.e. $\delta\, f^{(q)} \sim  f^{(q - 1)}$).
Finally, we note that  $\Delta\, f^{(q)} \sim  f^{(q)}$ which shows that the degree of a form remains 
intact when it is operated upon  by the Laplacian operator $\Delta$ of differential geometry.

In our present free $\mathcal{N} = 2$ SUSY theory, there is yet another physical
realization of the cohomological operators because we note that if we take a state 
$|\chi >_p$ which has an eigenvalue $p$ corresponding to   
$({\bar Q\,  Q}/{H})$, namely; 
\begin{eqnarray}
\left(\frac{\bar Q\,  Q}{H}\right)\,\left|\chi \right>_p = p\,  \left|\chi \right>_p, 
\end{eqnarray}
(where $p$ is a real  number), then, we have the following 
\begin{eqnarray}
&&\left(\frac{\bar Q\,  Q}{H}\right)\, \bar Q\,\left|\chi \right>_p 
= (p + 1)\,\bar Q\, \left|\chi \right>_p, \nonumber\\
&&\left(\frac{\bar Q\,  Q}{H}\right)\,  Q\,\left|\chi \right>_p 
= (p - 1)\, Q\, \left|\chi \right>_p, \nonumber\\
 && \left(\frac{\bar Q\,  Q}{H}\right)\, H\,\left|\chi \right>_p = p \, H\, \left|\psi \right>_p, 
\end{eqnarray}
which demonstrates that the states   $\bar Q\;|\chi >_p, \,  Q\;|\chi >_p$ and $H\;|\chi >_p$ have the eigenvalues 
$(p + 1), (p - 1)$ and  $p$, respectively.   This crucial observation  is {\it also} 
identical to the consequences that emerge after operation of the set
($d, \delta, \Delta$) on a differential form of degree $p$. Thus, we have the following mapping:  
 \begin{eqnarray}
(\bar Q,\;  Q, \;H)\,\Longleftrightarrow (d,\; \delta,\; \Delta).
\end{eqnarray} 
We conclude that, for our $\mathcal{N} = 2$ SUSY theory, there are {\it two} physical realizations of 
($d, \delta, \Delta$) in the language of conserved charges and their eigenvalues. 
If the degree of a given differential form is identified with the eigenvalue of a state 
in the total quantum Hilbert space of states (w.r.t. a specific hermitian operator),
 then, the operations of 
($d, \delta, \Delta$) on the above form exactly match with the operations of conserved charges 
on the specifically chosen quantum state of the theory in the Hilbert space. 

\section{$\mathcal{N} = 2$ SUSY symmetries: supervariable approach}

We can derive the nilpotent ($s_1^2 = 0, s_2^2 = 0$) symmetries $s_1$ and $s_2$ by using 
the supervariable approach [12]
where the SUSY invariant restrictions (SUSYIRs) play very important role. For the derivation of $s_1$
(cf. (3)), first of all, we generalize the dynamical variables ($z (t), \bar z(t), \psi(t), \bar\psi(t)$) 
to their counterparts
supervariables on the chiral supermanifold  ($Z(t, \theta), \bar Z(t, \theta), \Psi (t, \theta), \bar \Psi
(t, \theta)$) with the following expansions along the Grassmannian $\theta$-direction (see, e.g. [12]):  
\begin{eqnarray}
&&z(t) \; \longrightarrow \;  Z(t, \theta) = z(t) +  \theta\, f_1(t),\nonumber\\
 &&\bar z(t) \; \longrightarrow \; \bar Z(t, \theta) = \bar z(t) +  \theta\, f_2(t),\nonumber\\
&&\psi(t) \;\longrightarrow \; \Psi (t, \theta) = \psi (t)  + i\, \theta\, b_1 (t), \nonumber\\
&& \bar\psi (t)\;\longrightarrow \; \bar\Psi (t, \theta) = \bar\psi (t)  + i\, \theta\, b_2 (t),
\end{eqnarray} 
where the (1, 1)-dimensional chiral supermanifold is parametrized by ($t, \theta$)
and, as is evident, the secondary variables ($f_1, f_2 $) and ($b_1, b_2$) are  fermionic 
and bosonic in nature, respectively.

As has been pointed out in [12], the SUSYIRs require that the SUSY invariant quantities should
be independent of the ``soul" coordinates $\theta$ and $\bar\theta$ (with $\theta^2 = \bar\theta^2 = 0,\,
\theta\, \bar\theta + \bar\theta\, \theta$). For instance, it is clear (from (3)) 
that $s_1 \bar z = 0, s_1 \psi = 0$, which implies  that we have the following SUSYIRs:
\begin{eqnarray}
\bar Z (t, \theta) = \bar z (t), \qquad \Psi (t, \theta)
= \psi(t)\, \Longrightarrow \, f_2 = 0, \qquad b_1 = 0.
\end{eqnarray}
As a consequence, we have the $\theta$-independence of the chiral supervariables 
$\bar Z (t, \theta)$ and $\Psi (t, \theta)$. Now we note that $s_1\,(z^T \cdot \psi) = 0$ and
$s_1\, (\dot{\bar z} \cdot  z + \frac{i}{2}\, \bar\psi\cdot \psi) =0$. Thus, we have the following SUSYIRs
in our theory, namely;:
\begin{eqnarray}
&& Z^T (t, \theta)\cdot \Psi (t, \theta) = z^T (t)\cdot \psi (t), \nonumber\\
&& \dot{ \bar Z} (t, \theta) \cdot Z (t, \theta) + \frac{i}{2} \, \bar\Psi (t, \theta) 
\cdot \Psi (t, \theta)  = \dot {\bar z}(t)\cdot  z(t) + \frac{i}{2}\,\bar\psi(t)\cdot\psi(t),
\end{eqnarray}
where $z^T \cdot \psi = z_1 \psi_1 + z_2 \psi_2$. Taking the help of (24), it is clear that 
$f_1(t) \propto \psi (t)$ because of the top restriction in (25). 
Choosing $f_1 (t)= ({\psi (t)}/{\sqrt 2})$, we obtain
$b_2 (t) = ({2\,\dot {\bar z} (t)}/{\sqrt 2})$ from the bottom
restriction of (25). Plugging in these values in the expansions (23), we obtain
\begin{eqnarray}
&&Z^{(1)}(t, \theta) = z(t) +  \theta\, \left(\frac{\psi}{\sqrt 2} \right) \equiv z(t) + \theta \,(s_1\, z),\nonumber\\
&&{\bar Z}^{(1)}(t, \theta) = \bar z(t) +  \bar\theta\, (0) \equiv \bar z(t) 
+ \theta \,(s_1\, \bar z),\nonumber\\
&&\Psi^{(1)} (t, \theta) = \psi (t)  + \theta\,(0)
 \equiv \psi(t) + \theta\, (s_1\, \psi), \nonumber\\
&&\bar\Psi^{(1)} (t, \theta) = \bar\psi (t)  +  \theta \left(\frac{2 i \dot {\bar z}}{\sqrt 2} \right) 
\equiv \bar\psi (t)  + \theta (s_1 \bar\psi),
\end{eqnarray} 
where the superscript $(1)$ on the supervariables stands for the expansions of the chiral supervariables after application of the SUSYIRs (25).

A close look at (26) demonstrates that we  have already obtained the transformations $s_1$
(cf. (3)). Furthermore, we observe that the following mapping is true:
\begin{eqnarray}
\frac{\partial}{\partial \theta}\, \Big(\Omega^{(1)} (t, \theta) \Big) = s_1 \, \Omega (t),
\end{eqnarray} 
which establishes the connection between the translational generator $\partial_\theta$ on the chiral 
(1, 1)-dimensional supermanifold and the SUSY transformations $s_1$. In (27), $\Omega^{(1)} (t, \theta)$ is
 the generic supervariable obtained in (26) and $\Omega (t) = z, \bar z, \psi, \bar\psi$
denotes  the generic dynamical variable of the Lagrangian $L_0$ (cf. (2)). 

For the derivation of nilpotent ($s_2^2 = 0$) transformations $s_2$, first of all, 
we generalize the dynamical variables 
($z (t), \bar z(t), \psi(t), \bar\psi(t)$) to their counterparts
supervariables   ($Z(t, \bar\theta), \bar Z(t, \bar\theta), \Psi (t, \bar\theta), \bar \Psi(t, \bar\theta)$)
on the anti-chiral supermanifold with the following general expansions 
\begin{eqnarray}
&&z(t) \; \longrightarrow \;  Z(t, \bar\theta) = z(t) +  \bar\theta\, f_3(t),\nonumber\\
 &&\bar z(t) \; \longrightarrow \; \bar Z(t, \bar\theta) = \bar z(t) +  \bar\theta\, f_4(t),\nonumber\\
&&\psi(t) \;\longrightarrow \; \Psi (t, \bar\theta) = \psi (t)  + i\,\bar \theta\, b_3 (t), \nonumber\\
&& \bar\psi (t)\; \longrightarrow  \; \bar\Psi (t, \bar\theta) = \bar\psi (t)  + i\, \bar\theta\, b_4 (t),
\end{eqnarray} 
where the secondary variables ($f_3, f_4$) and ($b_3, b_4$) are fermionic and bosonic, respectively. We note that 
$s_2 z=0, s_2 \bar\psi = 0$. Thus, we have the following SUSYIRs: 
\begin{eqnarray}
Z (t, \bar\theta) = z (t), \qquad \bar\Psi (t, \bar \theta)
= \bar\psi(t)\, \Longrightarrow \, f_3 = 0, \qquad b_4 = 0.
\end{eqnarray}
To determine the other secondary variables ($f_4 (t), b_3 (t)$), we have the following SUSYIRs:
\begin{eqnarray}
&& \bar Z (t, \bar\theta)\cdot {\bar\Psi}^T (t, \bar\theta) = \bar z (t)\cdot {\bar\psi}^T (t), \nonumber\\
&& { \bar Z} (t, \bar\theta) \cdot \dot Z (t, \bar\theta) - \frac{i}{2} \, {\bar\Psi} (t, \bar\theta) 
\cdot \Psi (t, \bar\theta) 
  =  {\bar z}(t)\cdot  \dot z(t) - \frac{i}{2}\,{\bar\psi}(t)\cdot\psi(t),
\end{eqnarray}
because we note that the above expressions remain invariant $s_2 (\bar z \cdot {\bar\psi}^T) = 0,
s_2({\bar z}\cdot  \dot z - \frac{i}{2}\,{\bar\psi}\cdot\psi) = 0$ under  $s_2$.
Using inputs from (29), we obtain the following:
\begin{eqnarray}
f_4 (t) = \frac{\bar\psi (t)}{\sqrt 2}, \qquad\quad b_3 (t) = \frac{2\, \dot z (t)}{\sqrt 2}.
\end{eqnarray}
Thus, the expansions  (28) reduce to  
\begin{eqnarray}
&&Z^{(2)}(t, \bar\theta) = z(t) + \bar \theta\,(0) \equiv z(t) 
+ \bar\theta \,(s_2\, z),\nonumber\\
&&{\bar Z}^{(2)}(t, \bar\theta) = \bar z(t) +  \bar\theta\, \left(\frac{\bar\psi}{\sqrt 2} \right) \equiv \bar z(t) 
+ \bar\theta \,(s_2\, \bar z),\nonumber\\
&&\Psi^{(2)} (t, \bar\theta) = \psi (t)  + \bar\theta\,\left(\frac{2\,i\,\dot {\bar z}}{\sqrt 2} \right)
 \equiv \psi(t) + \bar\theta\, (s_2\, \psi), \nonumber\\
&&\bar\Psi^{(2)} (t, \bar\theta) = \bar\psi (t)  + \, \bar\theta\,(0)  
\equiv \bar\psi (t)  + \, \bar\theta\, (s_2\, \bar\psi),
\end{eqnarray} 
where the superscript $(2)$ denotes the expansions of the supervariables  after the application of the SUSYIRs 
(29) and (30). A careful observation of (32) 
demonstrates that we have already
derived  the SUSY transformations $s_2$ (cf. (3)). It is worth mentioning that 
the analogue of (27) can be defined for the SUSY transformations $s_2$ as well.

To provide the geometrical meaning to the nilpotency of the conserved charges $Q$ and $\bar Q$, we note 
the following 
\begin{eqnarray}
 Q &=&\frac{\partial}{\partial \theta}\, \Big[2\,\dot{\bar Z}^{(1)}(t, \theta)
\cdot Z^{(1)}(t, \theta)\Big]  
 \equiv  \frac{\partial}{\partial \theta}\, \Big[2\,\dot{\bar z}(t)
\cdot Z^{(1)}(t, \theta)\Big], \nonumber\\ 
 &=& \int d\theta\, \Big[2\,\dot{\bar Z}^{(1)}(t, \theta)
\cdot Z^{(1)}(t, \theta)\Big] \equiv \int d\theta\, \Big[2\,\dot{\bar z}(t)
\cdot Z^{(1)}(t, \theta)\Big],  \nonumber\\
Q &=& \frac{\partial}{\partial \theta}\, \Big[- i\,{\bar \Psi}^{(1)}(t, \theta)
\cdot \Psi^{(1)}(t, \theta)\Big]  
 \equiv \frac{\partial}{\partial \theta}\, \Big[- i\,{\bar \Psi}^{(1)}(t, \theta)
\cdot \psi(t)\Big], \nonumber\\ 
 &=& \int d\theta\, \Big[- i\,{\bar \Psi}^{(1)}(t, \theta)
\cdot \Psi^{(1)}(t, \theta)\Big]  
\equiv  \int d\theta\, \Big[- i\,{\bar \Psi}^{(1)}(t, \theta)
\cdot \psi(t)\Big],  \nonumber\\
\bar Q &=&\frac{\partial}{\partial \bar\theta}\, \Big[2\,{\bar Z}^{(2)}(t, \bar\theta)
\cdot {\dot Z}^{(2)}(t, \bar\theta)\Big] 
 \equiv  \frac{\partial}{\partial \bar\theta}\, \Big[2\,{\bar Z}^{(2)}(t, \bar\theta)
\cdot {\dot z}(t)\Big], \nonumber\\ 
&=& \int d \bar\theta\, \Big[2\,{\bar Z}^{(2)}(t, \bar\theta)
\cdot {\dot Z}^{(2)}(t, \bar\theta)\Big] 
 \equiv   \int d \bar\theta\, \Big[2\,{\bar Z}^{(2)}(t, \bar\theta)
\cdot {\dot z}(t)\Big],  \nonumber\\
 \bar Q &=&\frac{\partial}{\partial \bar\theta}\, \Big[+ i\,{\bar \Psi}^{(2)}(t, \bar\theta)
\cdot \Psi^{(2)}(t, \bar\theta)\Big]  
 \equiv  \frac{\partial}{\partial \bar\theta}\, \Big[+ i\,{\bar \psi}(t)
\cdot \Psi^{(2)}(t, \bar\theta)\Big], \nonumber\\ 
&=& \int d \bar\theta\, \Big[+ i\,{\bar \Psi}^{(2)}(t, \bar\theta)
\cdot \Psi^{(2)}(t, \bar\theta)\Big] 
 \equiv  \int d \bar\theta\, \Big[+ i\,{\bar \psi}(t)
\cdot \Psi^{(2)}(t, \bar\theta)\Big].
\end{eqnarray}
Thus, there are two different ways to express $Q$ and $\bar Q$ in the language of supervariables 
and Grassmannian derivatives. The nilpotency of $\partial_\theta$ and $\partial_{\bar\theta}$
(i.e. $\partial_\theta^2 = 0, \,\partial_{\bar\theta}^2 = 0$) implies 
that $\partial_\theta\, Q = 0, \partial_{\bar\theta}\, \bar Q = 0$. The latter imply expressions,  in the language of SUSY 
transformations $s_1$ and $s_2$
(and their generators) as: $s_1 Q = i\,\{ Q,\,  Q\} = 0$ and $s_2 \bar Q = i\,\{\bar Q,\, \bar Q\} = 0$ 
which prove the nilpotency ($Q^2 = \bar Q^2 = 0$) of the SUSY charges $Q$ and $\bar Q$. 
Furthermore, when we express the above expressions 
for $Q$ and $\bar Q$ (cf. (33)) in terms of the ordinary SUSY symmetries and dynamical 
variables, we observe that 
\begin{eqnarray}
  Q = s_1 \Big(2\, \dot{\bar z}\cdot z\Big) \equiv s_1\, \Big( - i\, \bar\psi \cdot\psi\Big), \qquad
 \bar Q = s_2 \Big(2\, {\bar z}\cdot \dot z\Big) \equiv s_2\, \Big( + i\, \bar\psi \cdot\psi\Big).
\end{eqnarray}
The above expressions   also prove the nilpotency of the conserved charges $Q$ and $\bar Q$ in a straightforward 
manner because $s_1^2 = 0, s_2^2 = 0$.

We can also capture the invariance of the Lagrangian $L_0$ (cf. (2)) in terms of the supervariables 
obtained after SUSYIRs. For instance, it can be checked explicitly that the following generalizations
of (2), namely;   
\begin{eqnarray}
&& L_0 \; \Longrightarrow \; {\tilde L}^{(ac)}_0 = 2\,\dot{\bar Z}^{(2)} \cdot \dot{ Z}^{(2)} 
 + \frac{i}{2}\, \left[\bar\Psi^{(2)} \cdot \dot{\Psi}^{(2)}  
- \dot{\bar\Psi}^{(2)} \cdot {\Psi}^{(2)}\right],\nonumber\\
&& L_0 \; \Longrightarrow \; {\tilde L}^{(c)}_0 = 2\,\dot{\bar Z}^{(1)} \cdot \dot{ Z}^{(1)} 
 + \frac{i}{2}\, \left[\bar\Psi^{(1)} \cdot \dot{\Psi}^{(1)} 
 - \dot{\bar\Psi}^{(1)} \cdot {\Psi}^{(1)}\right], 
\end{eqnarray}
(where the superscripts $(c)$ and $(ac)$ denote the chiral and anti-chiral nature of the Lagrangians 
${\tilde L}^{(c)}_0$  and ${\tilde L}^{(ac)}_0$, respectively) lead to  one of the key observations that
\begin{eqnarray}
\frac{\partial}{\partial\theta}\; \Big[{\tilde L}^{(c)}_0 \Big] =  s_1\, L_0
\equiv \frac{d}{dt}\Big(\frac{\dot{\bar z} \cdot \psi}{\sqrt 2} \Big), \nonumber\\
\frac{\partial}{\partial\bar\theta}\; \Big[{\tilde L}^{(ac)}_0 \Big] =  s_2\, L_0
\equiv \frac{d}{dt}\Big(\frac{\bar\psi\cdot {\dot z}}{\sqrt 2} \Big).
\end{eqnarray}
Geometrically, the invariance (cf. (4))  of the free Lagrangian is encoded in the above expressions (36).
It states that ${\tilde L}^{(c)}_0$  and ${\tilde L}^{(ac)}_0$ are the sum of  
composite supervariables  constructed from (26)
and (32) such that their translations along the $\theta$ and $\bar\theta$-directions of the 
(1, 1)-dimensional chiral and anti-chiral supermanifolds, respectively,  produce the  ordinary 
time-derivatives  in the ordinary 1D space thereby leading to symmetry invariance.

\section{Conclusions }

Our present endeavor is our first modest step towards our main goal of proving $\mathcal{N} = 2$ and 
$\mathcal{N} = 4$ interacting theories (of a charged particle constrained to move on a sphere in the background 
of a Dirac magnetic monopole) as tractable SUSY models for the Hodge theory. To achieve the above mentioned central goals, first of all, we have considered  the {\it free} $\mathcal{N} = 2$ version of the above interacting systems 
and established that it is a model for the Hodge theory.

In our present investigation, we have shown the physical realizations of the de Rham cohomological
operators in the language of symmetries (and conserved charges). We have also derived the $\mathcal{N} = 2$
nilpotent SUSY transformations by exploiting the supervariable approach [12] and provided geometrical 
meanings to them. In fact, as it turns out, the nilpotent $\mathcal{N} = 2$ SUSY transformations 
($s_1$ and $s_2$) are nothing but the translational generators ($\partial_\theta$ and $\partial_{\bar\theta}$)
along the $\theta$ and $\bar\theta$-directions of the chiral and anti-chiral supermanifolds on which the (0 + 1)-dimensional  dynamical   variables are generalized as supervariables. The nilpotency 
 of the transformations $s_1$ and $s_2$  are also encoded in such properties
   associated with  $\partial_\theta$ and $\partial_{\bar\theta}$.

 One of our immediate  goals is to prove that the interacting $\mathcal{N} = 2$ SUSY quantum mechanical model of a 
 charged particle, constrained to move on a sphere in the background of a Dirac magnetic monopole [20],
 is a tractable SUSY model for the Hodge theory. The most interesting  future endeavor for us is to find out the physical realizations of the cohomological operators in the case of $\mathcal{N} = 4$ SUSY quantum mechanical model [21] in the language of symmetry properties and  conserved charges. \\

\noindent
{\bf Acknowledgements} \\
One of us (RPM) would like to express his deep sense of gratitude to the Director and  members 
of THEP group at AS-ICTP for their cordial invitation  to visit the Centre where a major part of this
work was done. He also thankfully  acknowledges fruitful discussions with G. Thompson and M. Chaichian 
(Helsinki University).  SK gratefully acknowledges the financial support
from UGC, Government  of India, New Delhi, under its SRF-scheme.

\end{document}